\documentclass{ieeeaccess}
\usepackage{cite}
\usepackage{amsmath,amssymb,amsfonts}
\usepackage{algorithmic}
\usepackage{graphicx}
\usepackage{bm}
\usepackage{textcomp}
\usepackage{subfigure}
\usepackage[numbers,sort&compress]{natbib}
\usepackage{booktabs}
\usepackage[justification=centering]{caption} 
\def\BibTeX{{\rm B\kern-.05em{\sc i\kern-.025em b}\kern-.08em
    T\kern-.1667em\lower.7ex\hbox{E}\kern-.125emX}}
\begin{document}
\history{}
\doi{}

\title{Spread Mechanism and Influence Measurement of Online Rumors in China During the COVID-19 Pandemic}
\author{
\uppercase{Yiou Lin}, 
\uppercase{Hang Lei and Yu Deng}
}
\address{School of Information and Software Engineering, University of Electronic Science and Technology of China, Chengdu 610054, China}

\markboth
{Y Lin \headeretal: Spread Mechanism and Influence Measurement of Online Rumors in China During the COVID-19 Pandemic}
{Y Lin \headeretal: Spread Mechanism and Influence Measurement of Online Rumors in China During the COVID-19 Pandemic}

\corresp{Corresponding author:YIOU LIN. (e-mail: lyoshiwo@163.com).}

\begin{abstract}
In early 2020, the Corona Virus Disease 2019 (COVID-19) pandemic swept the world.
In China, COVID-19 has caused severe consequences. Moreover, online rumors during the COVID-19
pandemic increased people’s panic about public health and social stability. At present, understanding
and curbing the spread of online rumors is an urgent task. Therefore, we analyzed the rumor spreading
mechanism and propose a method to quantify a rumor’s influence by the speed of new insiders. The
search frequency of the rumor is used as an observation variable of new insiders. The peak coefficient
and the attenuation coefficient are calculated for the search frequency, which conforms to the exponential
distribution. We designed several rumor features and used the above two coefficients as predictable labels.
A 5-fold cross-validation experiment using the mean square error (MSE) as the loss function showed that the
decision tree was suitable for predicting the peak coefficient, and the linear regression model was ideal for
predicting the attenuation coefficient. Our feature analysis showed that precursor features were the most
important for the outbreak coefficient, while location information and rumor entity information were the
most important for the attenuation coefficient. Meanwhile, features that were conducive to the outbreak
were usually harmful to the continued spread of rumors. At the same time, anxiety was a crucial rumor causing factor. Finally, we discuss how to use deep learning technology to reduce the forecast loss by using
the Bidirectional Encoder Representations from Transformers (BERT) model.

\end{abstract}


\titlepgskip=-15pt

\maketitle

\section{Introduction}
\label{sec:introduction}
\IEEEPARstart{I}{n} December 2019, some hospitals in Wuhan, China
discovered multiple cases of unexplained pneumonia with
a history of exposure to the South China Seafood Market \cite{huang2020clinical}. The cases have now been confirmed as the acute respiratory infectious disease, named Corona Virus Disease 2019
(COVID-19). At the time of this writing (December 3, 2020), the global number of COVID-19 infections exceeded 64.7 million, and the deaths exceeded 1.5 million. The COVID-19 crisis has caused economic, social, and mental crises in a brief period, and it has spread internationally and affected all aspects of human life \cite{nicola2020socio}. When lockdowns and social distancing measures to prevent the spread of COVID-19 started, Wei $et\ al.$ pointed out that the prevalence of mental and psychological problems was as follows: sleep problems were 49.8\%, anxiety symptoms were 44.5\%. Among them, stress-induced excitatory symptoms accounted for 21.6\%, and depression symptoms were 18. 9\% \cite{wei2020psychological}. Meanwhile, the present COVID-19 pandemic appears to be leading to higher suicidality. Griffiths $et\ al.$  presented six cases in chronological order of couple suicides and attempted suicides relating to COVID-19 \cite{griffiths2020covid}. Research on rumors has a long tradition. For decades, one of the most popular ideas in rumor research is the idea that rumors are always accompanied by crises \cite{2016An}. According to a cross-sectional survey 30.9\% of all the respondents reported believing in some unverified COVID-19 crisis-related rumors from internet media \cite{cai2020novel}.
Now we are in an Internet era of information explosion. Recent theoretical developments have revealed that the role of the Internet and the media during the present pandemic crisis should not be underestimated \cite{pieri2019media}. Due to the lack of moderation and guarantee, as a platform
for freedom of speech, online social media and networks are
highly susceptible to the spread of rumors \cite{oh2013community}. 
For example,
after the Fukushima Nuclear Disaster in 2011, there was a
rumor that the nuclear leak would cause a short supply of salt. This rumor led to panic buying and hoarding salt. Similarly, misinformation about a COVID-19 medication regimen or unreliable protection measures has a severe impact on public health and can even affect social stability \cite{shabahang2020online}. For example, on January 31, 2020, it was reported that the Oral Liquid “Shuanghuanglian” (a Chinese patented medicine) had an inhibitory effect on the COVID-19. After the news was released, “Shuanghuanglian” products were quickly snapped up and hoarded, causing market chaos. Thus, quantitatively
describing rumor spreading and measuring the influence of 
rumors during the COVID-19 pandemic has a theoretical
significance. In the past several decades, the common strategies that have been used to study the rumor mechanism are simulation experiments and statistical learning. With the rise of deep learning, natural language processing (NLP) has been widely used in interdisciplinary fields \cite{liu2017survey}. This area of rumor
mechanism research constitutes a relatively new area that has emerged from NLP and deep learning technology.
\begin{table}[hbtp]
	\centering
	\caption{Chinese search engine market share in 2020}
	\begin{tabular}{lcllll}
		\toprule[2pt]
		\multicolumn{1}{c}{\textbf{Search Engine}} & Baidu      & Sogou                          & 360                        & Google                         & Bing                             \\
		\toprule
		\textbf{Share}                              & 69.55\% & \multicolumn{1}{c}{16.84\%} & \multicolumn{1}{c}{4.19\%} & \multicolumn{1}{c}{3.76\%} & \multicolumn{1}{c}{2.8\%}  \\
		\textbf{Period}                            & \multicolumn{5}{c}{From July 2019 to July 2020}\\
		\bottomrule[2pt]  	                                
	\end{tabular}
	\label{table0}
\end{table}

This study is based on a quantitative analysis of data extracted from Sogou (China's second-largest search engine company as shown in Table. \ref{table0}) search index. We believe that Internet surveillance is a convenient and cost-effective way to assess public responses and that it can provide evidence for government decision-making. The main contributions of this paper can be summarized as follows:
1. We propose an epidemic-like model to quantitatively describe the law of rumors spreading during the COVID pandemic. The model includes four-node states: susceptible state, infected state, refuted state, and removal state. Through
calculation by simulation, we found that, ideally, the density of the infected state conforms to a power-law distribution.

2. We collected a COVID-19 related rumor data crawled from search engine results based on which we propose a
method to determine the peak coefficient and the attenuation
coefficient of a rumor outbreak. As far as we know, no
previous research has investigated this.

3. Using NLP techniques and machine learning methods, we analyze the relationship between rumor features and rumor indexes, including the peak coefficient and the attenuation coefficient. Analysis of the experimental results reveal several exciting conclusions. We also ran deep learning techniques and showed how semantic features help improve peak coefficient prediction.

The remainder of this paper is organized as follows: In the next section, we first introduce the literature on rumor propagation and the NLP techniques used in this article. In the third section, we build the epidemic-like rumor model and describe the simulation experiments that were conducted. Then, we develop our hypotheses and the ideas for the verification of these hypotheses, and the peak coefficient and attenuation coefficient are calculated in the fourth section. After that, we discussed how three models predicated these two coefficients, their results, and analysis features. Finally, our conclusions are given and future research possibilities are suggested.

\section{Related works}
This section reviews several research papers and mentions a few indicative works about rumor theory and propagation. Rumor refers to a cooperative transaction in which community members offer, evaluate, and interpret information to reach a common understanding of uncertain situations, alleviate social tension, and solve collective crisis problems \cite{bordia1997rumor,bordia1999rumor,bordia2004problem}. From its birth, a rumor involves communication dynamics surrounding shared issues in a community, the generation and transmission are inseparable in practice \cite{oh2013community}. Rumor propagation is often modeled as a process of social contagion \cite{kawachi2008rumor}; therefore, epidemiological models are highly relevant to the research literature on rumor propagation \cite{daley2001epidemic}. Kermack and McKendrick studied the mathematical theory of epidemics and proposed the Susceptible-Infected-Removed (SIR) model, which is the most widely used model in research of rumor spreading \cite{kermack1927contribution}. Currently, many studies have proposed more rumor spreading models through the addition and modification of node state types and state transformation mechanisms \cite{yin2020covid,hui2020spread}. We propose a rumor spread
mechanism motivated by the above work. In addition to the
above simulation research, empirical research of rumors has
attracted people’s attention. Better understanding the spread of rumors on social networks (such as Twitter and Weibo) is especially valued \cite{2016An,li2020social}. However, using rumor forwarding in social networks as the rumor spread metric has two main disadvantages: 1. The users of social networks are limited and consist of mainly young people. 2. The dissemination data becomes unreliable when rumor information is gradually deleted. In response to this situation, the search index of web search engines has become a common area of focus. Search index data has been applied in many ways, such as forecasting tourist arrivals \cite{sun2019forecasting,volchek2019forecasting}, hotel registrations \cite{rivera2016dynamic}, economic indicators \cite{mclaren2011using}, and monitoring influenza epidemics \cite{yuan2013monitoring}. The search index is also helpful for epidemic prevention and control. An extensive search trends-based analysis of public attention proved that Internet monitoring could be particularly incredibly and economical in the prevention and control of epidemics and rumors \cite{xie2020extensive}. Meanwhile, social media is also helping to predict the number of COVID-2019 cases \cite{qin2020prediction}, map of health literacy and social panic \cite{xu2020mapping}, and predict the peak of the COVID-19 pandemic through Internet search \cite{effenberger2020association}.
Although the search index research has been widely used, its application in combination with the field of rumor analysis has not yet been proposed. Therefore, this article attempts to unearth the inner connection between user searches and online rumors. 

 From a modeling perspective, we need to quantify user searches and online rumors separately. The user’s search behavior leaves a search record and forms a search index. Thus, the real difficulty lies in how to express online rumors. Online rumors can be regarded as a kind of short text, and the short text representation method is very mature \cite{tang2012enriching,qazvinian2011rumor}. In this study, we employed content-based features as well as semantic features. The technologies we used included part of speech (POS) \cite{ratnaparkhi1996maximum}, named entity recognition (NER) \cite{lample2016neural}, and Bidirectional Encoder Representations from Transformers (BERT) \cite{devlin2018bert}.

The internal connection between two variables is a mathematical mapping relationship, which can be described by a machine learning model. By combining machine learning and the representation of rumors, Raveena $et\ al.$ investigated in retrospect a dataset in which tweets spreading rumor detection was completed machine learning algorithms, including the k-nearest neighbor and Naive Bayes classifier \cite{dayani2015rumor}. The first to propose the application of rumor theory to social media and community intelligence was Onook Oh $et\ al.$, who studied citizen-driven information processing \cite{oh2013community}. The above research is limited to the identification of the authenticity of rumors. Li $et\ al.$ analyzed the association between user features and rumor forwarding behavior in five main rumor categories: economics, society, disaster, politics, and military \cite{li2020social}. However, their work failed to consider the influence of rumor
features on forwarding behavior. Therefore, due to these
shortcomings, our method is proposed to quantify the rumor
spread influence based on the search frequency and rumor
representation.

\section{Spread Mechanism}

After rumors appear in real social networks, the government or relevant authoritative organizations dispel the rumors through mainstream media or the Internet. However, existing research on rumor dissemination mostly involves the modeling of single rumor without refuting information \cite{nekovee2007theory,xia2015rumor} or only analyzed the key factors affecting mass communication behavior were analyzed from the information dissemination level \cite{zhao2016analysis}. To more reasonably describe the interactive propagation process and to understand the interactive propagation mechanism of different rumors, this study used mean-field theory,
which is commonly used in complex network propagation
dynamics, to establish equations to characterize the dynamics
propagation model of the interactive process. The impact
of rumor rejection nodes is discussed and the key factors
affecting the spread of rumors based on the simulation results
are analyzed below.

Drawing lessons from the spreading characteristics of viruses in a network, the attitudes of users on rumors and other information in social networks are combined to classify
the status of users in the network.

\begin{figure}[htbp]
	\centering
	\includegraphics[scale=0.19]{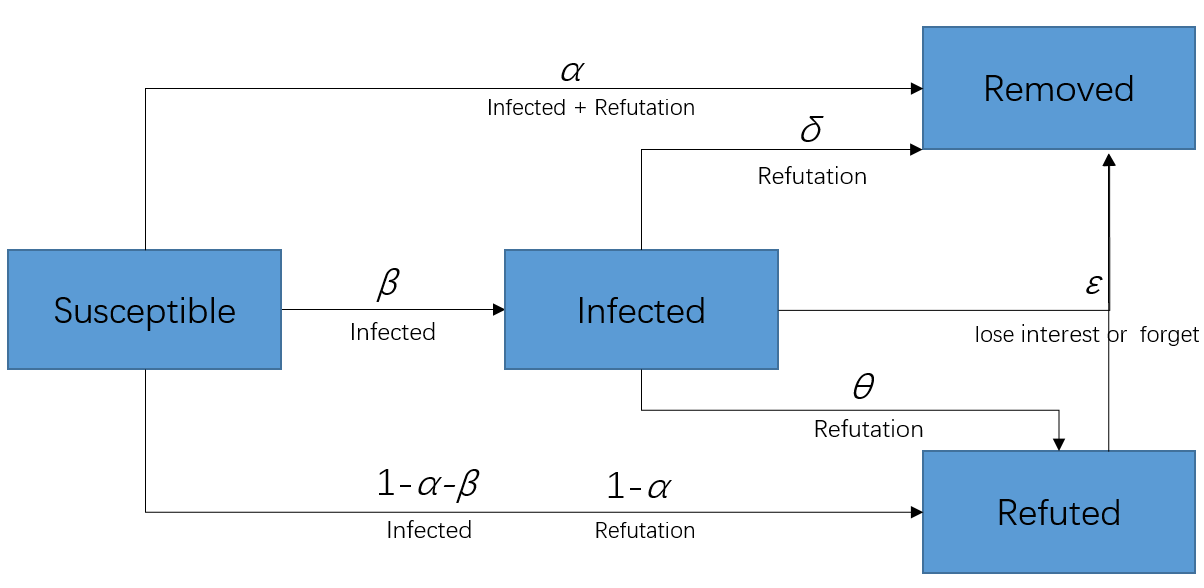}
	\caption{Flow chart of rumor spreading with four states}
	\label{Fig1}
\end{figure}

In the process of rumor spreading, there are four node states: susceptible state, infected state, refuted state, and removal state. Users who are infected by refuted node are the disseminators of positive information, including those who publish positive information; users who believe in authoritative information will disseminate it. Removal nodes say that such users will neither be infected by rumors nor by positive information. All the users in the network are regarded as nodes, and the relationship between friends is regarded as edges. The transition relationship between node states as shown in Fig. \ref{Fig1}.

In Fig. \ref{Fig1}, susceptible means those who are ignorant, Infected means those who will be a disseminator, removed means those who lost interest in the rumor, and refuted means those who see through the rumor and disseminate positive information. The status change is based on the following four rules.

1) If a susceptible node contacts a rumor-infected node, then it will change to a rumor-infected state with a probability of $\beta$, change to removed state with a probability of $\alpha$, and change to a refuted node with a probability of $1-\beta-\alpha$.  

2) If a susceptible node contacts a refuted node,  then it will change to a removed state with a probability of $\alpha$ or change to another refuted node with a probability of $1-\alpha$.

3) If an infected node contacts a refuted node, then it will change to removed state with a probability of $\sigma$, change to a refuted node with a probability of $\theta$, or remain unchanged.

4) In each iteration, an infected node or a refuted node has the probability of $\varepsilon$ of being forgotten and changes to be a removed node.

\begin{figure}[htbp]
	\centering
	\includegraphics[scale=0.45]{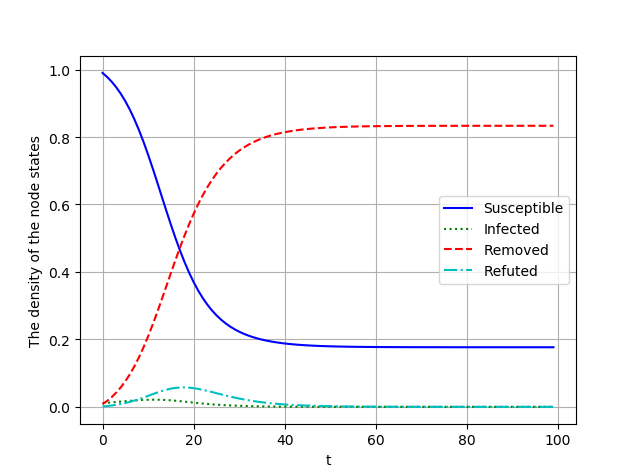}
	\caption{State density change with time}
	\label{Fig2}
\end{figure}

In a homogeneous network, $S(t)$, $I(t)$, $R_l(t)$, and $R_2(t)$ denote the density of a susceptible node, infected node, removed node (who know about the rumor but do not care about it), and refuted node (who do not accept the rumor and try to stop it) through the rumors at time t, respectively. They reach the normalization condition $S(t) + I(t) + R_1(t) + R_2(t) = 1$. The above dissemination rules can be combined to establish the average spread of rumors, and the mean-field equations can be described as follows:

\begin{equation}\label{E1}
\begin{aligned}
\frac{\mathrm{d} S(t)}{\mathrm{d} t}=&-\alpha\langle k\rangle (I(t) +R_2(t))S(t)-\beta\langle k \rangle I(t)S(t)\\
&-\langle k\rangle S(t)((1-\alpha-\beta)I(t)+(1-\alpha)R_2(t))\\
=&-\langle k\rangle (I(t) +R_2(t))S(t),
\end{aligned}
\end{equation}
\begin{equation}\label{E2}
\begin{aligned}
&\frac{\mathrm{d} I(t)}{\mathrm{d} t} = (\beta\langle k \rangle S(t) -(\delta+\theta)\langle k\rangle R_2(t)-\epsilon)I(t),
\end{aligned}
\end{equation}
\begin{equation}\label{E3}
\begin{aligned}
\frac{\mathrm{d} R_1(t)}{\mathrm{d} t}= &\alpha\langle k\rangle (I(t) +R_2(t))S(t)+\delta\langle k\rangle I(t)R_2(t)\\&+\epsilon(I(t)+R_2(t)),
\end{aligned}
\end{equation}
and
\begin{equation}\label{E4}
\begin{aligned}
\frac{\mathrm{d} R_2(t)}{\mathrm{d} t}=&\langle k\rangle S(t)((1-\alpha-\beta)I(t)+(1-\alpha)R_2(t))\\&+\theta\langle k\rangle I(t)R_2(t)-\epsilon R_2(t),
\end{aligned}
\end{equation}

where $\langle k\rangle$ is the average degree of the network, which was usually set to 1 in the simulation experiment. There are few spreaders at the beginning of the rumor spreading; thus, it is usually assumed that there is only a very small number of infected nodes in the initial state network, and the rest are marked as susceptible nodes. Afterward, the number of the infected nodes first increases to the peak and then decreases until it goes down to zero, at which point the rumor dies out, and the system reaches a stable state.  
The various node densities $S(t),I(t), R_1(t)$ and $R_2(t)$ under these four states evolve with time t (see Fig. \ref{Fig2}). The parameters used in the numerical simulation are $N=10000,\alpha=0.6,\beta=0.3,\delta=0.1,\epsilon=0.2,\theta=0.3,I(0)=0.01$. 

Figure \ref{Fig2} shows that the spread of rumors caused about 80\% of the population to be infected by rumors in the end. Because of the forgetting rate $\epsilon$, people lose interest in rumors or their factors cease to spread and eventually become removed.  At this time, record the density of new insiders who knew but did not spread the rumors at the end is recorded as $R$, and then it can be known that $R=\operatorname{final}\{R(t)\}=\underset {t \rightarrow +\infty}{\lim} R(t)=R(\infty)=R_2(+\infty)=1-S(+\infty)\approx-\int_{0}^{+\infty} S^{\prime}(t) \mathrm{d}t$. 

\begin{figure}[htbp]
	\centering
	\includegraphics[scale=0.45]{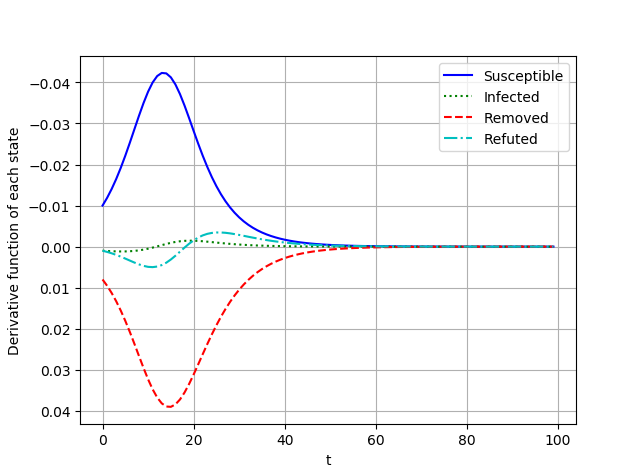}
	\caption{Derivative function of each state with time}
	\label{Fig3}
\end{figure}

We use the search frequency of a rumor as an observation variable of new insiders. Search frequency was first used to track public events, such as influenza-like illnesses, by Google in 2009 \cite{GINSBERG2009Detecting}. Experiments proved that the frequency of specific queries was positively correlated with relative events. The size of $S^{\prime}(t)$ represents the rate of increase of people who are new to the rumor, and those who are new to the rumors will use the Internet to query the rumors with a fixed probability and leave a query record. It can be found that $S^{\prime}(t)$ conforms to a power-law distribution, as shown in Fig \ref{Fig3}. The same conclusion and case studies can be found in rumor spreading among micro-blog followers based on user browsing behavior\cite{6602630}. Ideally, assuming that susceptible persons will query the rumors with the same probability after being exposed to them, then the query index of the rumors should have a similar distribution. Like a roller coaster track, the rate of spread of rumors will first accelerate and increase, and then it will decelerate and decrease. The rate of new contacts of rumors rises to the apex first, then falls, and the rate of decline gradually slows down. Furthermore, the rate has an apparent long-tail distribution.

\begin{figure*}[htbp]
	\centering
	\includegraphics[scale=0.45]{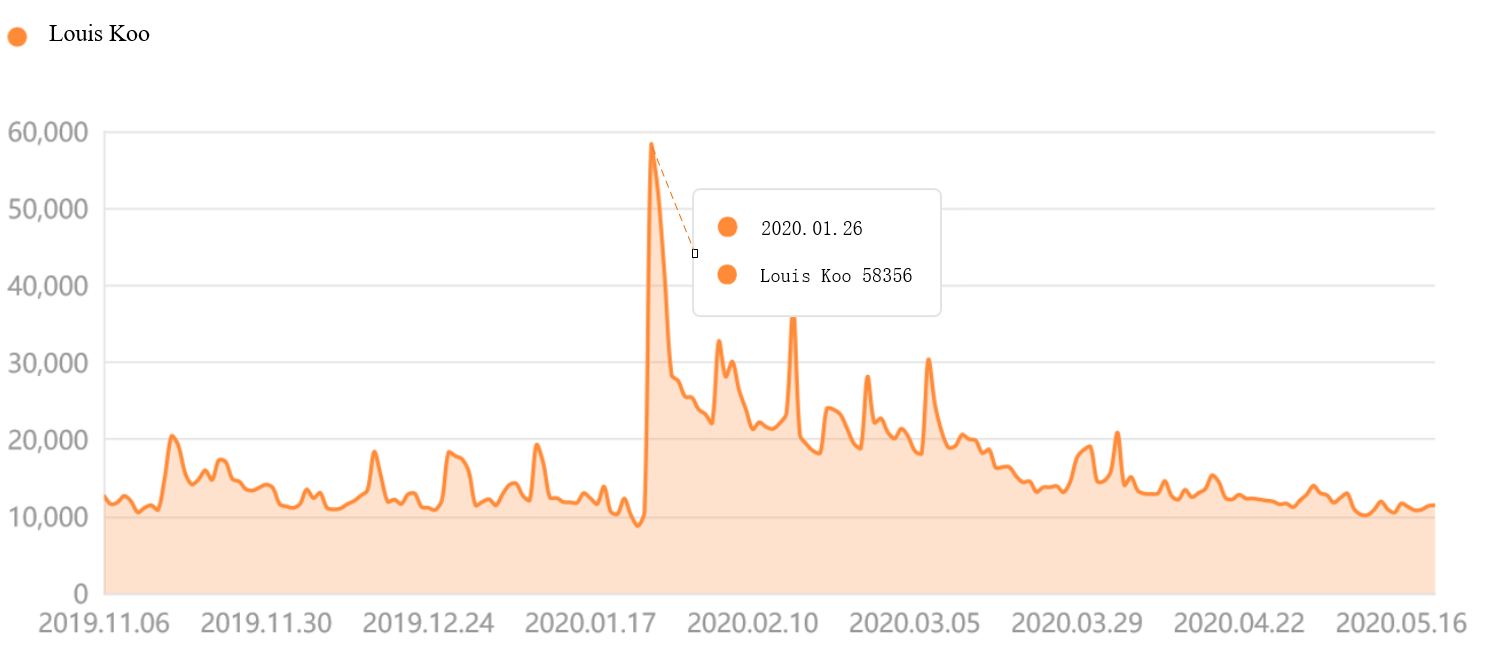}
	\caption{Louis Koo's search frequency provided by Sogou}
	\label{Fig4}
\end{figure*}
\begin{figure}[htbp]
	\centering
	\includegraphics[scale=0.2]{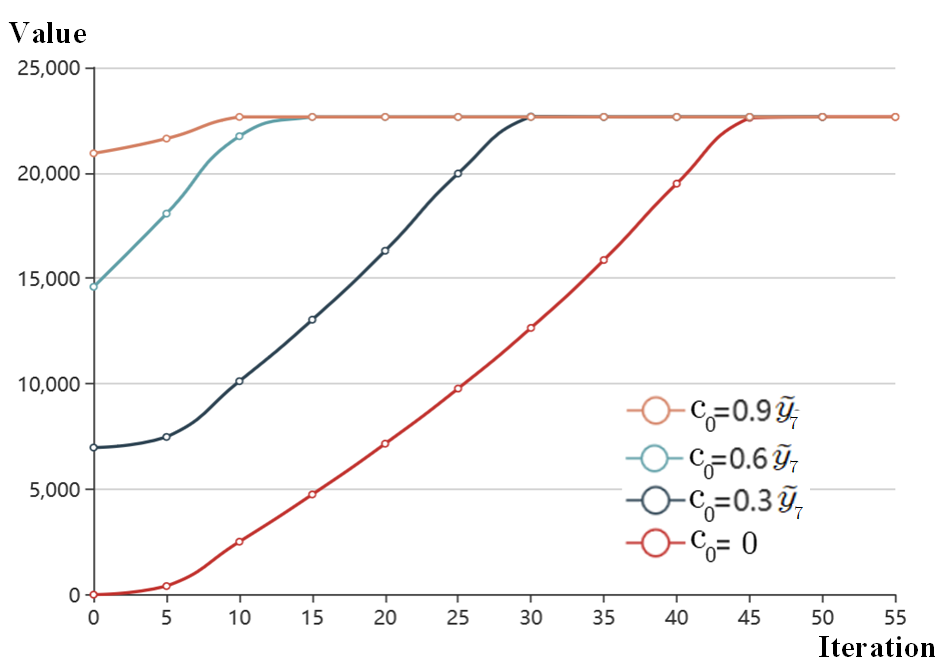}
	\caption{Convergence for parameter $c$}
	\label{Figc}
\end{figure}
\section{Rumour modeling}
\subsection{Model simulation}
Compared with the rate of spread of rumors based on the spread mechanism, we found that the actual result was slightly different. As an example, the search index of the rumor "Louis Koo donated 10 million yuan to Wuhan" provided by Sogou indicates is shown in Fig. \ref{Fig4}. Louis Koo received unusual attention when the rumor occurred (January 26, 2020), and it reached its highest within 24 hours. Compared with Fig. \ref{Fig3}, We suppose that the reasons for the above phenomenon are as follows: In an era when traffic is king, in order to attract attention, news media and self-media always forward rumors as soon as possible. The right to speak a public opinion is in the hands of news media and self-media. Rumors are pushed directly to interested users with push
technology and hot searches. The rumor also caused Louis
Koo-related searches to increase. After nearly two months, the search value of Louis Koo's search index returned to the historical average.
Since rumors usually peak on the first day, they fit the exponential decline distribution more than the power-law distribution. 
\begin{equation}\label{E5}
\begin{aligned}
y_{t}=e^{at+b}+c
\end{aligned}
\end{equation}
Mark the search frequency motivated by rumor as $\tilde{y}_{\mathrm{t}}=y_{\mathrm{t}}-c$ and 
$\mathbf{Y}=\left[\ln \left(\mathrm{y}_{0}-c\right), \ln \left(\mathrm{y}_{1}-c\right), \ldots, \ln \left(\mathrm{y}_{n}-c\right)\right]^{\mathrm{T}}$
where $t$ is the time, $n$ represents the total number of days, $a$ and $b$ are rumor-related variables, $c$ is the bias which is unrelated to rumor (initialized with the minimum search frequency multiply by 0.9 in seven days), and $y$ is the search frequency.
\begin{equation}\label{E6}
\begin{aligned}
J_{L S}(\theta)=\frac{1}{2}\|\mathbf{X} \theta-\mathbf{Y}\|^{2} \cdot 
\end{aligned}
\end{equation}
$\text { s.t. } \quad \mathbf{X}=\left(\begin{array}{cccc}
1 &1&\dots&1 \\
0 &1&\dots&n

\end{array}\right)^ \text{T} \text { and } \theta=(a, b)$,
thus let the gradient of $J_{L S}(\theta)$ be $\mathbf{0}$, we can get
\begin{equation}\label{E7}
\begin{aligned}
\nabla_{\theta} J_{L S}=\left(\frac{\partial J_{L S}}{\partial \theta_{1}}, \frac{\partial J_{L S}}{\partial \theta_{2}}\right)=\mathbf{X}^{\mathrm{T}} \mathbf{X} \theta-\mathbf{X}^{\mathrm{T}} \mathbf{Y}=\mathbf{0}
\end{aligned}
\end{equation}
\begin{equation}\label{E8}
\begin{aligned}
\theta=(a, b)=\left(\mathbf{X}^{\mathrm{T}} \mathbf{X}\right)^{-1} \mathbf{X}^{\mathrm{T}} \mathbf{Y}
\end{aligned}
\end{equation}
\begin{equation}\label{E9}
\begin{aligned}
c=\left(\sum_{0}^{n} y_{t}-\sum_{0}^{n} \tilde{y}_{t}\right) / \mathrm{n}
\end{aligned}
\end{equation}

Equations (\ref{E6})-(\ref{E9}) are calculated iteratively several times until parameter $c$ convergence. Using the least square method, Fig. \ref{Figc} shows the variation of parameter $c$ with the number of iterations under different initial conditions. Our observation
found that usually the intensity of a rumor will not exceed seven days and can be calculated by rumor-related variables $a$ and $b$. For each rumor, parameter $a$ is the attenuation coefficient, and $b$ is the peak coefficient of each rumor.
Since $\tilde{y}_{\mathrm{t}}$ is a geometric series, the sum of $\tilde{y}_{\mathrm{t}}$ could be calculated as follows:
\begin{equation}\label{E10}
\begin{aligned}
\sum_{0}^{n} \tilde{y}_{t}=&\tilde{y}_{0} \cdot\left(1-e^{a(n+1)}\right) /\left(1-e^{a}\right)\\=&\mathrm{e}^{b} \cdot\left(1-e^{a(n+1)}\right) /\left(1-e^{a}\right).
\end{aligned}
\end{equation}
 For the rumor "Louis Koo donated 10 million yuan to Wuhan" the solution $a$, $b$, and $\tilde{c}$ were finally equal to -0.605, 10.69 and 22653 respectively. After experiments, the range of $a$ was determined to be [-0.26,-1.45], and then the following inequation could be obtained:
\begin{equation}\label{E11}
\begin{aligned}
1.31 \tilde{y}_{0} \leq \sum_{0}^{+\infty} \tilde{y}_{t}=\tilde{y}_{0} /\left(1-e^{a}\right) \leq 4.37 \tilde{y}_{0}
\end{aligned}
\end{equation}
From what has been discussed above, usually the intensity of a rumor will not exceed five times the search frequency of the first outbreak day, and it can be expressed by the sum of the search frequency.

\begin{figure*}[htbp]
\centering
\includegraphics[scale=0.35]{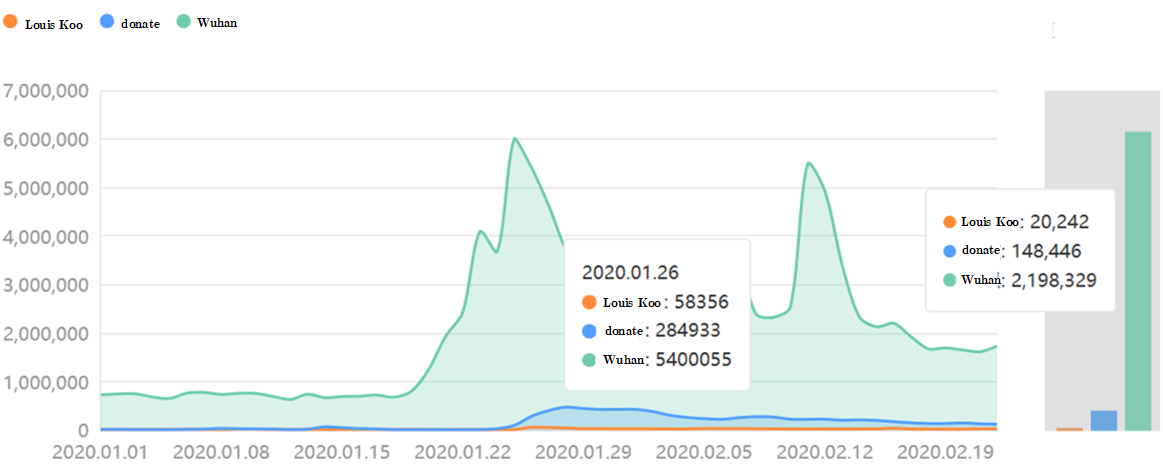}
\caption{Search frequency trends of key named entities in the Louis Koo rumor}
\label{Fig6}
\end{figure*}
\subsection{model hypotheses}
In the following sections, our goal is to study the elements that can be used to estimate the rumor-related variables $a$ and $b$. For this goal, we introduce five empirically testable hypotheses.

1.	The fundamental entity that owns the least search frequency can represent the rumor search index.

2.	Public anxiety helped spread the rumor.

3.	The outbreak intensity is associated with historical awareness of named entities of each rumor.

4.	The daily search frequency sequence before the rumor can be used to predict the burst strength of the rumor.

5.	Under the background of the pandemic, the feedback of network media is associated with rumor spread.

In this work, Hypothesis 1 is the premise of our research. As shown in Fig. \ref{Fig6}, other common entities usually had far more searches than the fundamental entity, which means they are not representative. Public sentiment was calculated by the daily average sentiment of Weibo text, as shown in Fig. \ref{Fig8}. The historical awareness of named entities was measured by natural logarithms of the average daily search frequency in the last two months of 2019. The feedback of the network media was presented by the results returned by the search engine, as shown in Fig. \ref{Fig7}.
\section{Experiment and result analysis}

\subsection{Dataset and Feature extraction}
\begin{figure}[htbp]
	\centering
	\includegraphics[scale=0.35]{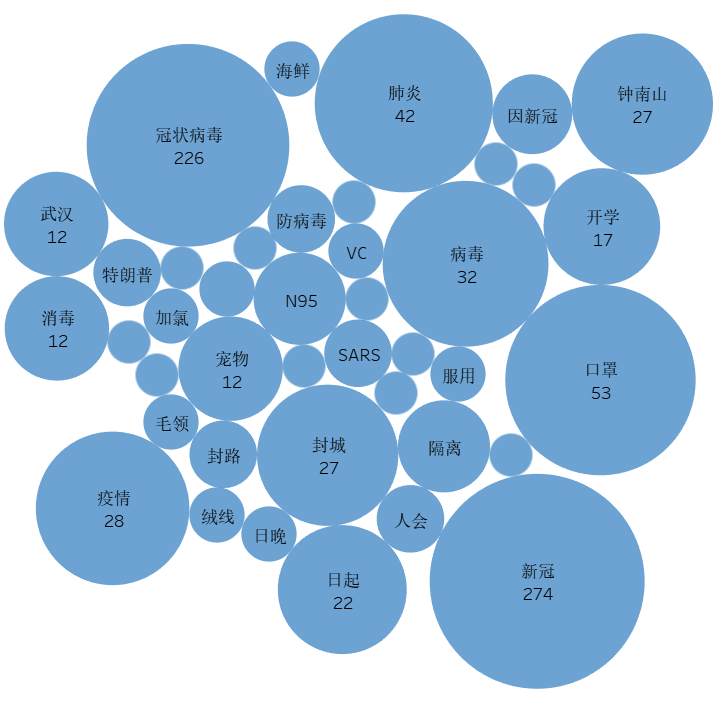}
	\caption{Common rumor keyword cloud}
	\label{cloud}
\end{figure}
 We use crawler frame selection to collect a public rumor corpus including 1029 rumors from Dingxiangyuan (a medical knowledge-sharing website) and Tencent (an Internet-based platform company). For each rumor, we extracted two keywords and drew the word cloud graph that is shown in Fig. \ref{cloud}. Rumors abounded with some elements, such as viruses, drugs, celebrities, protections, and locations. Since the search frequency of commonly named entities (such as “Wuhan” and “donate”) were easily influenced by other rumors or facts, the fundamental entity (“Louis Koo”) that had the least search frequency had obvious guiding significance, as shown in Fig. \ref{Fig4} and Fig. \ref{Fig6}. We empirically accept Hypothesis 1 and manually filtered 120 rumor based on it, of which variable $a$ and $b$ could be accurately estimated and parameterized. 
To find the key features that could accurately predict variable $a$ and $b$ of each type of situational information,
we extracted the Boolean features of six NER tags.
 The historical awareness of the top two most commonly named entities and the fundamental entity before the outbreak of the pandemic were selected. Meanwhile, a rumor portrait was obtained through the search engine. An example is shown in Fig.\ref{Fig7} of the search results about the rumor "Louis Koo donated 10 million yuan to Wuhan". Three elements were given attention, including the resulting amount (224,000), start date (January 26, 2020), and rumor flag (fake news). The resulting amount was used as statistical feature feedback by network media.

\begin{figure}[htbp]
	\centering
	\includegraphics[scale=0.45]{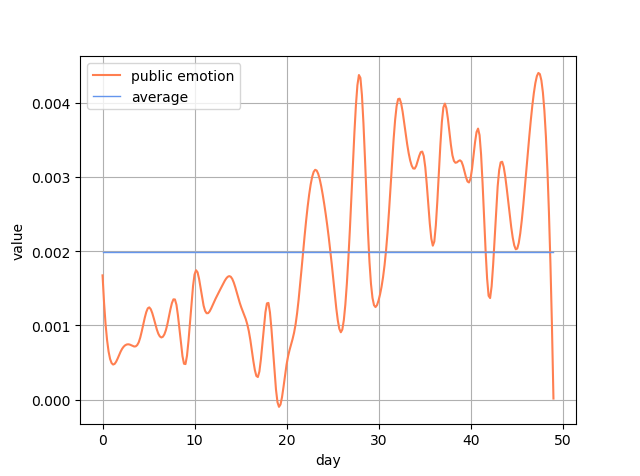}
	\caption{An emotional time distribution map of Internet users since January 1, 2020}
	\label{Fig8}
\end{figure}

\begin{figure}[htbp]
\centering
\includegraphics[scale=0.45]{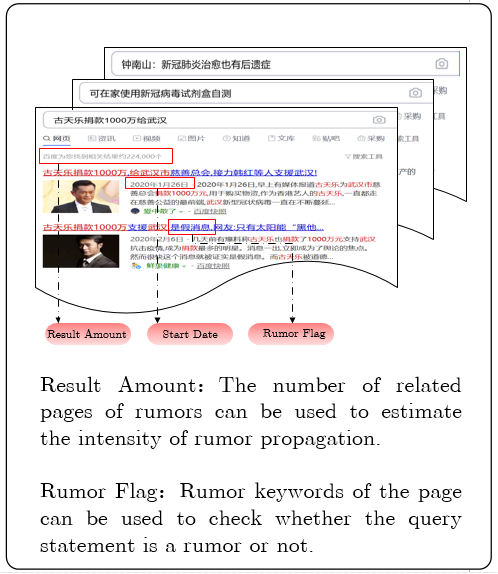}
\caption{Results returned by search engines that used rumors as the queries}
\label{Fig7}
\end{figure}

 Allport and Postman \cite{jack2018humanitarian} proposed the idea that a rumor is motivated by intellectual pressure along with the emotional. Moreover, Anthony \cite{Anthony1973Anxiety} introduced anxiety as a proxy variable for rumor-mongering conditions. Thus, to Hypothesis 2, We used a public COVID-19 related microblog corpus\footnote{https://www.datafountain.cn/competitions/423}. As shown in Fig. \ref{Fig8}, the index of public sentiment reached its lowest point before Wuhan was closed due to the pandemic. After that, the index rose quickly and fluctuated violently. Using the start data extracted from the search results of rumor, as shown in Fig. \ref{Fig8}, the current public emotion density value could be located. Meanwhile, to verify Hypothesis 3, the search frequency sequence of the fundamental named entity in a week before the rumor breakout is extracted.

In summary, a rumor like "Louis Koo donated 10 million yuan to Wuhan" could be extracted into [(top2: "donated", ln(800)), (top1: "Wuhan", ln(2,200,000)), (key: "Louis Koo", ln(11,000)), (PER, 1), (ORG, 0), (LAC, 1), (NZ, 0), (N, 0), (V, 1), ("resulting amount", ln(224,000)), ("public emotion", 0.0032), ("search frequency sequence before the rumor broke", (ln(13850), ln(10584), ln(10278), ln(12281), ln(10105), ln(8738), ln(10548)))]. At last, all the features were transformed into the range [0, 1] using Min-Max Normalization.

\subsection{Regression models}
Machine learning methods, such as the linear regression model and the decision tree model, have been applied in the field of rumor analysis \cite{oh2013community,2020Social}. The linear regression model is one of the most basic types of statistical techniques and is widely used in predictive analysis. It shows a relationship between two variables with a linear algorithm and equation. The decision tree model is a decision support tool that uses a tree-like model of decisions and their possible consequences. The main advantage of the decision tree classifier is its ability to use different feature subsets while still being readable. After using min-max normalization to covert features to the same scale, a linear regression model measures the significance of features by weights directly, while a tree model can handle nonlinear situations. The mean-square error (MSE) is used as a loss function and Formula (\ref{E6})-(\ref{E8}) calculates linear weights. The decision tree learning includes three steps: feature selection, decision tree generation, and decision tree pruning. To build the tree, we used the CART algorithm to train a regression tree.
Compared to the linear model using 5-folds cross-validation, experiments have shown that the decision tree model reduced the MSE of variable $b$ from 3.77 to 1.79 but increased the MSE of variable $a$ from 0.14 to 0.21. We extracted the essential features for each model, and they are listed in Table.\ref{table1}.

In the next section, we discuss the significant differences in the use of features by regression models, including the linear regression model and decision tree model.

\begin{table*}[hbtp]
\centering
\caption{Most important features for linear regression model and decision tree model}
\begin{tabular}{@{}clcc@{}}
	\toprule[2pt]
	&Description & Weight for $a$ & Importance for $b$ \\ 
	\toprule
	PER & The Boolean variable that indicates whether or not the person's name is present & $1.8*10^{-1}$&0\\ 
	ORG & The Boolean variable that indicates whether or not the organization is present & $-1.7*10^{-1}$&0\\ 
	LOC & The Boolean variable that indicates whether or not the location is present & $\bm{2.2*10^{-1}}$&$1.3*10^{-2}$\\ 
	Top-1 & Historical search frequency of the most commonly named entity & $-1.7*10^{-1}$&$1.0*10^{-2}$\\ 
	SFR-1 & Search frequency of the fundamental named entity at the day before the rumor break out & $-1.2*10^{-1}$&$\bm{6.2*10^{-1}}$\\ 
	PE & Public emotion when rumor occurs & $-9*10^{-2}$&$1.2*10^{-1}$\\
	ANE & Average search frequency of the fundamental named entity & $2.1*10^{-1}$&$1.8*10^{-1}$\\ 
	RA & Resulting amount of the search engine & $1.6*10^{-1}$&$2.6*10^{-1}$\\ 
	\bottomrule[2pt]
\end{tabular}
\label{table1}
\end{table*}
\subsection{feature analysis}
Through the experiment, we found that only three kinds of named entities were useful in the prediction of the attenuation coefficient (parameter $a$), including PER, ORG and LOC, but only LOC made sense for the peak coefficient (parameter $b$). This suggests that rumors involving institutions are more quickly quelled because the authorities will deny it in time, while rumors involving specific celebrities and locations are more likely to be believed by people. Two other conclusions can be drawn. One is that Top-1, SFR-1, and PE were not conducive to the persistence of rumors, while ANE and RA were opposites. Second, all five of these features were essential to predicting parameter $b$ and SFR-1 was the most important.

Since most rumors die down within seven days, one possible inference is that the feature that makes rumors more intense is not conducive to their sustainable renewal. A correlation analysis of features and parameters is shown below using the Pearson correlation coefficient. Table. \ref{table2} shows that this inference was not entirely correct because ANE and RA were positively correlated with all three parameters $a,b,c$.
\begin{table}[hbtp]
\centering
\caption{Pearson correlation coefficient between six features and three parameters}
\begin{tabular}{@{}cccc@{}}
	\toprule[2pt]
	&coefficient for $a$&coefficient for $b$&coefficient for $c$\\ \toprule
	LOC & 0.27 &-0.16&0.22\\
	Top-1&	-0.15&	0.10&	-0.27\\ 
	SFR-1&	-0.04&	0.58&	0.50\\
	PE&	-0.24&	0.02&	0.07\\
	ANE&	0.15&	0.41&	0.57\\
	RA&	0.09&	0.42&	0.18\\
	\bottomrule[2pt]
\end{tabular}
\label{table2}
\end{table}
Regarding Hypothesis 2, PE and parameter $a$ were negatively correlated, that is to say, when the public mood is negative, the rumor search frequency dropped more slowly, and the rumor is to spread more continuously. At the same time, the outbreak of rumors was also related to PE. Specifically, for the tree model, the importance of PE for parameter $b$ is $1.2*10^{-1}$, and the linear correlation of $b$ was only 0.02, indicating that the outbreak degree of the rumor and PE was not a simple linear relationship. Hypothesis 3 was also correct, but only ANE and TOP-1 played a significant role in entity search records. Hypothesis 4 can be refined to "the outbreak of rumors could be predicted by precursors". The experiments showed that only the precursor that happened the day before the rumor made sense due to the suddenness of the rumor's explosion. Compared with other features, SFR-1 (the intensity of precursors) had the strongest positive correlation with the intensity of the rumor outbreak. In terms of Hypothesis 5, we only used RA as the feedback of the search engine in the statistical characteristics, and the results showed that rumors with a more massive RA tended to have a stronger breakout and spread for a longer amount of time.
\subsection{Deep and Wide model}
\begin{figure}[hbtp]
	\centering
	\includegraphics[scale=0.16]{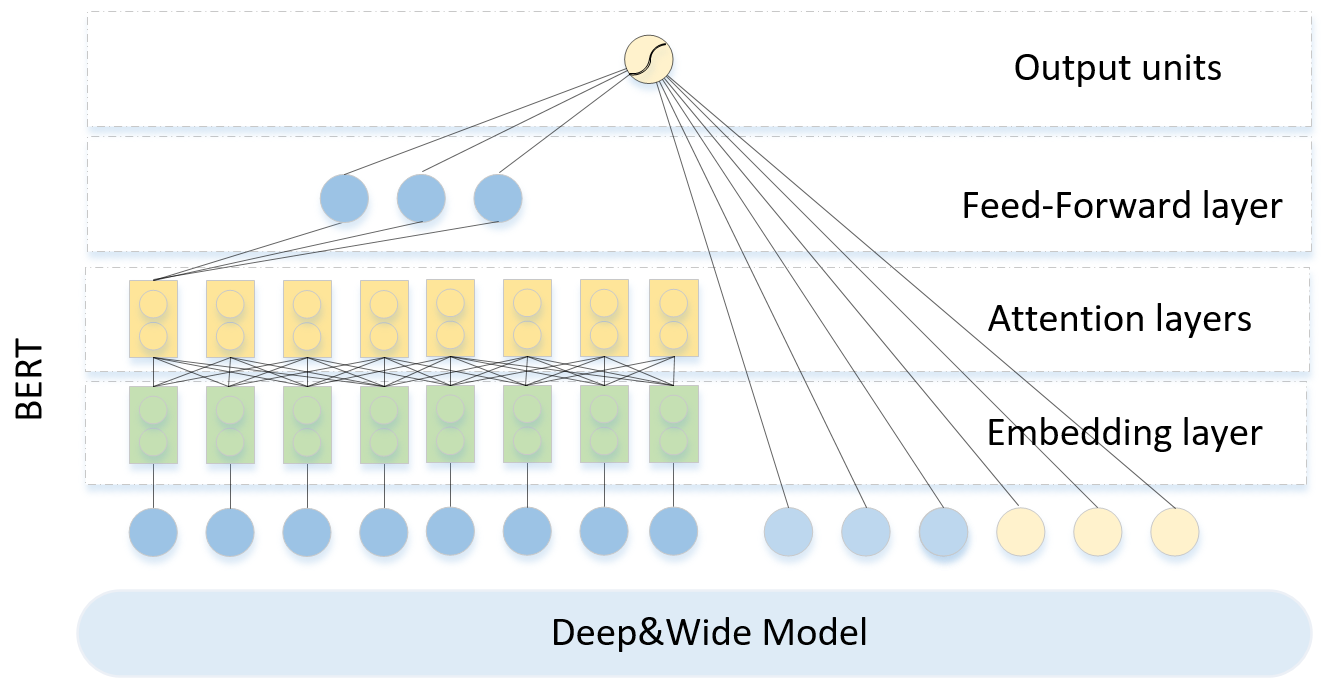}
	\caption{The spectrum of Deep and Wide model}
	\label{Fig9}
\end{figure}
In the field of NLP, BERT has apparent advantages in small sample learning and transfer learning. We used both the rumor data and the result text on the first page of the search engine as the training corpus. 

We propose a hybrid model that applies the common deep and wide framework of the recommendation system to solve the task. As the deep part is directly implemented using the Chinese BERT, we only introduce statistical features of the wide part in detail. The statistical features include the eight features shown in Table. \ref{table1}.

\begin{figure*} 
	\centering 
	\subfigure[ Loss of model using original corpus ]{
		 \label{fig:subfig:a}
		  \includegraphics[width=3in]{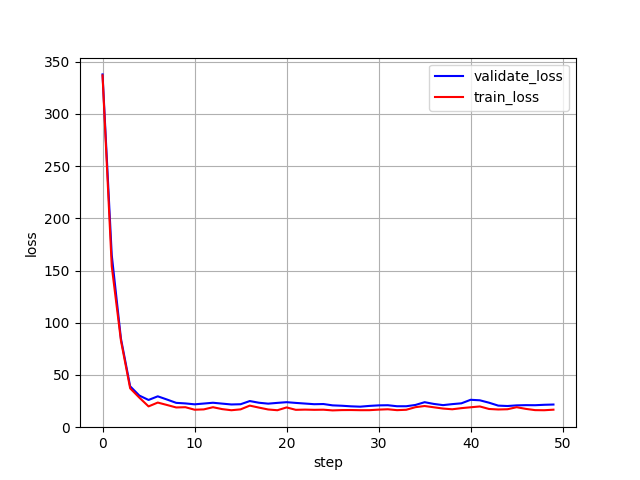}} 
	  \hspace{0.3in} \subfigure [Loss of model using clean corpus]{
		 \label{fig:subfig:b}
		  \includegraphics[width=3in]{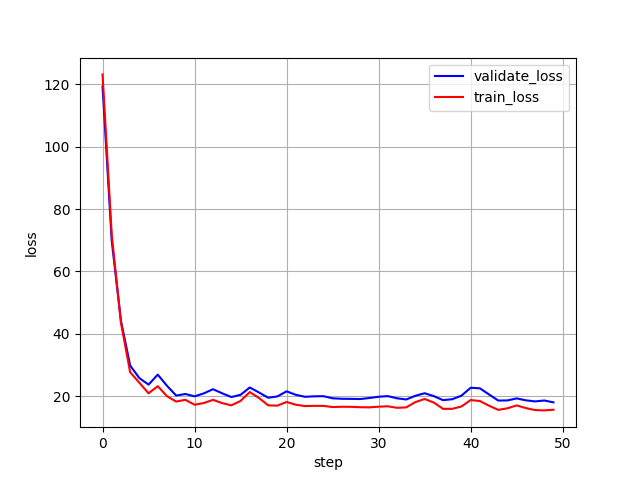}}
		 \caption{The loss changes with the number of epoch} 
		\label{fig-double} 
	\end{figure*}

As shown in Fig. \ref{Fig7}, the search engine returned not only the RA but also the text content of the first page. Using BERT encoding, the semantic vectors with 256 dimensions could be obtained. However, the vector dimension exceeded the number of samples, causing an overfitting problem. Thus, we did not modify the original weight of the BERT model as a solution to avoid overfitting; instead, a transfer learning method was used. The text contents of the front pages of all the rumors were collected and compressed into three dimensions using the hidden layers of three neurons to predict the RA. We used 50\% of the data as the training set and 50\% of the data as the validation set. As most of the information on the first page is redundant, it only consists of repeated news or debunking of the rumor. Thus, the key differentials are the rumor itself, the release time, and the publishers. Experiments show that this operation helps the model converge preferably with fewer than two points MAE of RA, and the results are shown in Fig. \ref{fig-double}.
We speculate that the timely authoritative refutation of rumor sources plays a perfect role in hindering the spread of a rumor. At the same time, this implies that the outbreak of rumor is strong enough to attract the attention of those involved in refuting the rumor. By using deep and wide model, the MSE of parameter $a$ was reduced from 0.14 to 0.10. Meanwhile, the output of the three semantic hidden layer neurons extracted by the deep and wide model was added to the tree model, and the experiment showed that the MSE of parameter $b$ increased from 1.79 to 1.83. This means that information on the front page barely predicts the outbreak intensity. Table. \ref{table3} shows that the first two semantic features made sense in promoting the prediction result of parameter $a$. We assume that semantic features do not work for parameter $b$ because it duplicates the function of the RA.
\begin{table}[hbtp]
\centering
\caption{ Correlation coefficient and importance between semantic represents and parameters}
\begin{tabular}{@{}cccc@{}}
	\toprule[2pt]
	&coefficient for  $a$&coefficient for $b$&importance of $c$\\ \toprule
	Neure-1 & -0.16 &0.15&$9*10^{-2}$\\
	Neure-2&	-0.28&	0.05&	0\\ 
	Neure-3&	-0.01&0.38&	$1.2*10^{-1}$\\
	RA&	0.09&	0.42&	$2.6*10^{-1}$\\
	\bottomrule[2pt]
\end{tabular}
\label{table3}
\end{table}
\section{Conclusion and AND FUTURE WORK}
We constructed a corpus of rumors in China during the pandemic and proposed a method to determine the peak coefficient and attenuation coefficient of rumor outbreak based on the change of the fundamental entity search index. Based on this method, we calculated the corresponding indexes of rumors and analyze the relationship between the characteristics and the indexes. In the future, we will continue to study the differences and connections between rumors and the propagation behaviors on ordinary hot events on the Internet. Moreover, we will study the use of unified models to describe the parameters. More data should be used and a unified model method should be trained to identify the peak coefficient and attenuation coefficient of rumor outbreaks. The current method still requires filtering the training set manually and is highly dependent on feature engineering. In the future, we plan to apply automatic labeling methods and the language model to avoid this limitation.

\bibliographystyle{IEEEtran}
\bibliography{ref}

\begin{IEEEbiography}[{\includegraphics[width=1in,height=1.25in,clip,keepaspectratio]{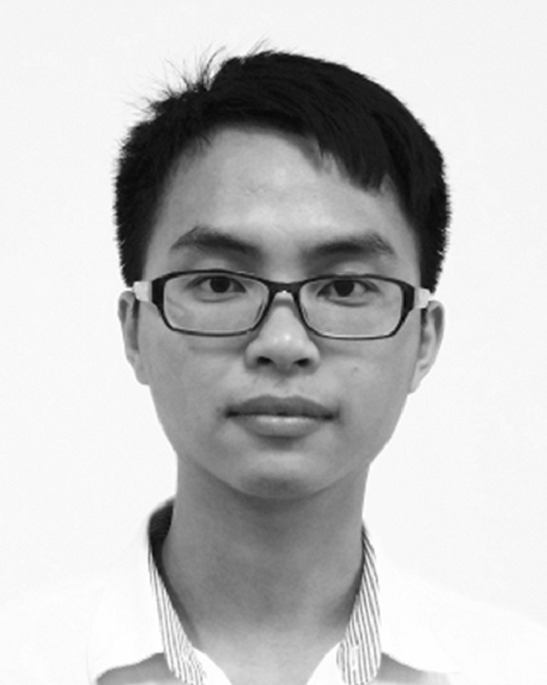}}]{YIOU LIN} received the bachelor's degree from the School of Information and Software Engineering, University of Electronic Science and Technology of China, China in 2013, where he is currently pursuing the Ph.D. degree. His research interests include deep learning and natural language processing.
\end{IEEEbiography}

\begin{IEEEbiography}[{\includegraphics[width=1in,height=1.25in,clip,keepaspectratio]{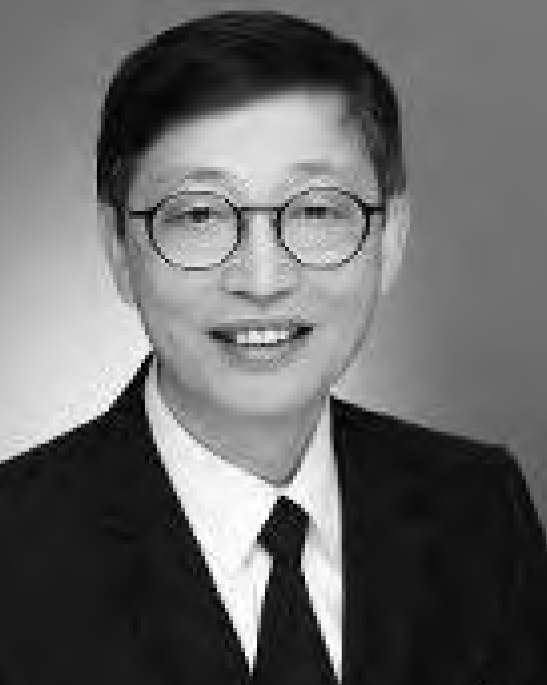}}]{HANG LEI} received the Ph.D. degree in computer science from the University of Electronic Science and Technology of China, China, in 1997. After graduation, he conducted research in the fields of real-time embedded operating systems, operating system security, and program verification as a Professor with the Department of Computer Science, University of Electronic Science and Technology of China, where he is currently a Professor (a Doctoral Supervisor) with the School of Information and Software Engineering. His research interests include big data analytics, machine learning, and program verification.
	
\end{IEEEbiography}

\begin{IEEEbiography}[{\includegraphics[width=1in,height=1.25in,clip,keepaspectratio]{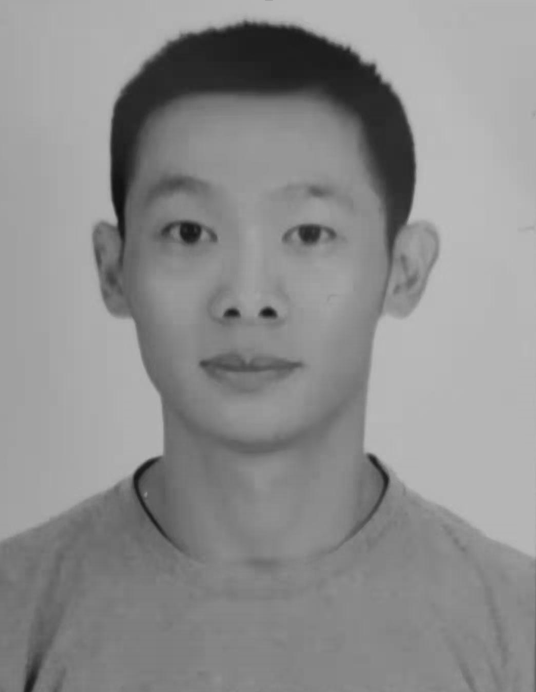}}]{YU DENG} has been pursuing a Ph.D. degree in the School of Information and Software Engineering, University of Electronic Science and Technology of China since 2014. His research interests include deep learning and sentiment analysis.
\end{IEEEbiography}

\EOD

\end{document}